\documentclass[twocolumn,showpacs,preprintnumbers,amsmath,amssymb]{revtex4}
\usepackage{graphicx}  
\usepackage{dcolumn}   
\usepackage{bm}        
\usepackage{amssymb}   
\usepackage{array}
\usepackage[colorlinks,citecolor=black,linkcolor=black,hyperindex,pdfborder={0 0 1},dvipdfm]{hyperref}%

\begin{document}
\title{Quantum resource studied from the perspective of quantum state superposition
}

\author{Chengjun Wu,$^1$ Junhui Li,$^1$ Bin Luo,$^2$ and Hong Guo$^{1,~}$}
\email[Correspondence author:\ ]{hongguo@pku.edu.cn}

\affiliation{$^1$State Key Laboratory of Advanced Optical Communication,
Systems and Networks, School of Electronics Engineering and Computer
Science, and Center for Quantum Information Technology, Peking
University, Beijing 100871, China}

\affiliation{$^2$State Key Laboratory of Information Photonics and Optical Communications, Beijing University of Posts and Telecommunications, Beijing 100876, China}

\date{\today}

\begin{abstract}
Quantum resources,such as discord and entanglement, are crucial in quantum information  processing. In this paper, quantum resources are studied from the aspect of quantum  state superposition.
We define the local superposition (LS)
as the superposition between basis of single part, and nonlocal
superposition (NLS) as the superposition between product basis of
multiple parts.
For quantum resource with nonzero LS, quantum operation must be introduced to prepare it, and for quantum resource with nonzero NLS, nonlocal quantum operation must be introduced to prepare it. We prove that LS vanishes if and only if the state is classical and NLS vanishes if and only if the state is separable. From this superposition aspect, quantum resources are categorized as superpositions existing in different parts. These results are helpful to study  quantum resources from a unified frame.

\end{abstract}




\maketitle

\section{Introduction}
Quantum resources are crucial in quantum information processing
\cite{quantum information, QERMP, QDRMP, GQI}. Quantum entanglement,
the first proposed quantum resource, has shown to be important in
quantum computation \cite{quantum information, QERMP, QDRMP},
quantum teleportation \cite{Teleportation, TeleportationE},
superdense coding \cite{superdense coding}, quantum cryptography
\cite{QKD1, QCWS}, Bell test \cite{testing Bell, MVBellCM}, etc.
However, states without quantum entanglement but with nonzero
quantum discord \cite{quantum discord} are also useful in some
quantum information process, e.g., deterministic quantum computation
with one qubit \cite{DQC1, DQC1E}, remote quantum state preparation
\cite{RSP1, RSPNP}, thus also being considered as quantum resource.
These resources have been studied from several perspectives, such as
Von Neumann entropy\cite{EOF, concurrence,quantum discord},
distance\cite{relative entropy of entanglement,unified
REB,geometrical discord,unified GD}, and so on\cite{suddenDeath,
Schmidt rank, MID, SEnQN}. Since all entangled states have nonzero
discord, entangled states are considered as a subset of quantum
discordant states\cite{quantum discord}, but whether entanglement
and discord are different resources is still an open question.

In this paper, we study the quantum resources from the aspect of quantum state superposition.
For a state in composite system, we define the local superposition
(LS) as the superposition between basis of single part, and
nonlocal superposition (NLS) as the superposition between product
basis of multiple parts. In this state superposition framework, quantum resources can be categorized by the superposition between different kinds of basis including different parts. We show that a state is classical when and only when it has zero local superposition and a
state is separable when and only when it has zero nonlocal
superposition. As for the relation of state superposition to the previously proposed quantum entanglement and quantum discord, quantum entanglement can be considered as the quantification of nonlocal superposition but excluding the local superposition, while
quantum discord includes the
local superposition. However, whether quantum discord also includes nonlocal superposition is not determined in this paper.

Moreover, the LS and NLS introduced hereinafter is generalized to
multidimensional and multipartite systems. For multipartite states,
we find that for pure states with Schmidt decomposition, NLS equals
LS, while for pure states without Schmidt decomposition, NLS and LS
might be different. This state superposition perspective might be
useful in studying the multipartite entanglement. For multipartite
systems, we show explicitly the different kinds of state
superpositions existing in different single part of the system, some
composite parts, or all parts of the system. These plural
superpositions, thus plural resources, provide us with many ways to
explore the quantum feature of quantum systems.

This paper is organized as follows. In part II, we first introduce
the local superposition in two state bipartite system and then
generalize it to multidimensional bipartite system. In part III,
nonlocal superposition for two partite system is introduced. In part
IV, LS and NLS are defined and analyzed for multipartite system. In
the V part, several discussions are given. In part VI, main
conclusions are presented. In the appendix, some proofs of the
statements  in the paper are provided.

\section{Local superposition in bipartite system}
When considering a quantum state with
superposition in the $\left\{ {\left| 0 \right\rangle ,\left| 1
\right\rangle } \right\}$ basis, $\left| \psi \right\rangle  = \alpha
\left| 0 \right\rangle  + \beta \left| 1 \right\rangle $, the
coherence of the state can be defined as $C(\left\{ {\left| 0
\right\rangle ,\left| 1 \right\rangle } \right\}) = 2\left| \alpha
\right|\left| \beta  \right| = 2{\left( {P(\left| 0 \right\rangle
)P(\left| 1 \right\rangle )} \right)^{1/2}}$, where ${\left| \alpha
\right|^2} = P(\left| 0 \right\rangle ) = \left\langle {\psi }
 \mathrel{\left | {\vphantom {\psi  0}}
 \right. \kern-\nulldelimiterspace}
 {0} \right\rangle \left\langle {0}
 \mathrel{\left | {\vphantom {0 \psi }}
 \right. \kern-\nulldelimiterspace}
 {\psi } \right\rangle $, ${\left| \beta  \right|^2} = P(\left| 1 \right\rangle ) = \left\langle {\psi }
 \mathrel{\left | {\vphantom {\psi  1}}
 \right. \kern-\nulldelimiterspace}
 {1} \right\rangle \left\langle {1}
 \mathrel{\left | {\vphantom {1 \psi }}
 \right. \kern-\nulldelimiterspace}
 {\psi } \right\rangle $, and ${P(\left| 0 \right\rangle )}$ and ${P(\left| 1 \right\rangle )}$
 are the probability for state $\left| \psi  \right\rangle $ in state
 $\left| 0 \right\rangle $ and $\left| 1 \right\rangle $, respectively.
 Zero coherence indicates that there is no
state superposition, and thus this state has no superposition
between basis $\left\{ {\left| 0 \right\rangle ,\left| 1
\right\rangle } \right\}$. On the contrary, if the coherence gets
the maximum value $1$, there is perfect state superposition
between basis $\left\{ {\left| 0 \right\rangle ,\left| 1
\right\rangle } \right\}$. There are other cases between these two
extreme ones. Therefore, it is reasonable to define the amount of
superposition of $\left| \psi \right\rangle $ between basis
$\left\{ {\left| 0 \right\rangle ,\left| 1 \right\rangle }
\right\}$ as
\begin{equation}\label{eq1}
\begin{split}
S(\left| \psi  \right\rangle ,\left\{ {\left| 0 \right\rangle ,\left| 1 \right\rangle } \right\}) = 2{\left( {P(\left| 0 \right\rangle )P(\left| 1 \right\rangle )} \right)^{1/2}}. \\
\end{split}
\end{equation}
Eq. (\ref{eq1}) indicates that when $S(\left| \psi  \right\rangle
,\left\{ {\left| 0 \right\rangle ,\left| 1 \right\rangle }
\right\})\ne 0$, there must be state superposition between ${\left|
0 \right\rangle }$ and ${\left| 1 \right\rangle }$. Note that the value in Eq. (\ref{eq1}) is dependent on the basis chosen, that is the amount of superposition defined above is basis specific.

Consider a system A in a composite system AB, for example, a
singlet state, $\left| {{\psi ^ - }} \right\rangle  =
\frac{1}{{\sqrt 2 }}\left( {\left| {01} \right\rangle  - \left|
{10} \right\rangle } \right)$, no matter which
basis for system A is used, it is hard to say that A is in a state
superposition of that basis. Actually, in this case, it is
impossible to assign a state description of system A. However, for
this state, when the basis $\left\{ {\left| 0 \right\rangle ,\left| 1
\right\rangle } \right\}$ is chosen, system A will behave exactly as
if it is in state $\frac{1}{{\sqrt 2 }}\left( {\left| {0}
\right\rangle  - \left| {1} \right\rangle } \right)$, assuming
that there is no operation on B. (Note that, this assumption is
true, since if there is some operation on B, then this composite
system should be described by a different state.) Thus, from this
point, we generalize the state superposition in system A and
define the amount of superposition of this state in system A in
the same way as Eq. (\ref{eq1}). Generally, for a two qubits pure
state ${\left| \psi \right\rangle _{AB}}$, we define
the amount of superposition of A in the $\left\{ {{{\left|
\varphi \right\rangle }_A},{{\left| {{\varphi ^ \bot }}
\right\rangle }_A}} \right\}$ basis (through this paper, when basis is
mentioned, it refers to orthonormal basis) as
\begin{equation}\label{eq2}
\begin{split}
&S_A(\left| \psi  \right\rangle {  _{AB}},\left\{ {{{\left| \varphi  \right\rangle }_A},{{\left| {{\varphi ^ \bot }} \right\rangle }_A}} \right\}) = \\
&2{{\left( {P({{\left| \varphi  \right\rangle }_A})P({{\left| {{\varphi ^ \bot }} \right\rangle }_A})} \right)}^{1/2}} , \\
 \end{split}
\end{equation}
where $P({\left| \varphi  \right\rangle _A}) = {\rm Tr}_B\left[
{{}_A{{\left\langle {\varphi }
 \mathrel{\left | {\vphantom {\varphi  \psi }}
 \right. \kern-\nulldelimiterspace}
 {\psi } \right\rangle }_{AB}}{{\left\langle {\psi }
 \mathrel{\left | {\vphantom {\psi  \varphi }}
 \right. \kern-\nulldelimiterspace}
 {\varphi } \right\rangle }_A}} \right]$, which is the probability for system A in state ${\left| \varphi  \right\rangle _A}$.
 More generally, for a specific decomposition of mixed state $\left\{ {{p_i},\;{{\left| {{\psi _i}} \right\rangle }_{AB}}} \right\}$
(${\rho _{AB}} = \sum\limits_i {{p_i}{{\left| {{\psi _i}}
\right\rangle }_{AB}}\left\langle {{\psi _i}} \right|} $), we define
the amount of superposition of A in $\left\{ {{{\left| \varphi
\right\rangle }_A},{{\left| {{\varphi ^ \bot }} \right\rangle }_A}}
\right\}$ basis as
\begin{equation}\nonumber
\begin{split}
\sum\limits_i {{p_i}S_A({{\left| {{\psi _i}} \right\rangle }_{AB}},
 \left\{ {{{\left| \varphi  \right\rangle }_A},{{\left| {{\varphi ^ \bot }} \right\rangle }_A}} \right\})} . \\
\end{split}
\end{equation}
The above superposition is introduced by the basis of part A, and we
define this as the local superposition with respect to part A.

\emph{\textbf{Definition: }}The
measure of LS of ${\rho _{AB}}$ for part A is defined as
\begin{equation}\label{eq3}
\begin{split}
  LS_A(\rho ) = \min \sum\limits_i {{p_i}S_A({\left| {{\psi _i}} \right\rangle } _{AB},\left\{ {{{\left| \varphi  \right\rangle }_A},
  {{\left| {{\varphi ^ \bot }} \right\rangle }_A}} \right\})}  ,\\
\end{split}
\end{equation}
where the minimum is taken over all decompositions $\left\{
{{p_i},\;{{\left| {{\psi _i}} \right\rangle }_{AB}}} \right\}$ and
all basis $\left\{ {{{\left| \varphi  \right\rangle }_A},{{\left|
{{\varphi ^ \bot }} \right\rangle }_A}} \right\}$.

Note that the LS defined in Eq.
(\ref{eq3}) is basis independent.  From the perspective of preparation, we have the choice to choose different basis. For $LS_A = 0$, there exists a
certain decomposition and basis that no superposition presents,
and thus part A can be prepared via classical operation (we use classical operation to mean the operation that no quantum superposition is introduced and use quantum operation to mean the other). For $LS_A
\ne 0$, no matter which decomposition and basis are considered,
there is superposition in part A, and thus part A can only be
prepared via quantum operation. For example: for state ${\left| \psi  \right\rangle _{AB}} = \frac{1}{2}\left( {{{\left| 0 \right\rangle }_A} + {{\left| 1 \right\rangle }_A}} \right)\left( {{{\left| 0 \right\rangle }_B} + {{\left| 1 \right\rangle }_B}} \right)$, the local superposition in the $\left\{ {{{\left| 0 \right\rangle }_A},{{\left| 1 \right\rangle }_A}} \right\}$ basis is nonzero and quantum operation is needed for the preparation when this basis is chosen, while the local superposition between basis $\left\{ {\frac{1}{{\sqrt 2 }}\left( {{{\left| 0 \right\rangle }_A} + {{\left| 1 \right\rangle }_A}} \right),\frac{1}{{\sqrt 2 }}\left( {{{\left| 0 \right\rangle }_A} - {{\left| 1 \right\rangle }_A}} \right)} \right\}$ is zero and classical operation is enough for the preparation when this basis is chosen.
Thus, from the aspect of preparation, Eq.
(\ref{eq3}) shows the minimum amount of superposition
produced in part A when preparing this state.

\emph{\textbf{Theorem 1:}} For pure state ${\left| \psi  \right\rangle _{AB}}$, $LS_A = 0$ iff the state is a product state.

Proof. For general
pure state ${\left| \psi  \right\rangle _{AB}} = {a_{00}}\left| {00}
\right\rangle  + {a_{01}}\left| {01} \right\rangle  + {a_{10}}\left|
{10} \right\rangle  + {a_{11}}\left| {11} \right\rangle$, by Schmidt decomposition, it can be written as ${\left| \psi  \right\rangle _{AB}} = \alpha
\left| {0'0'} \right\rangle  + \beta \left| {1'1'} \right\rangle $, where
$\alpha$ and $\beta$ are the singular value of matrix $A$
(with ${a_{ij}}$ being its elements) and $\alpha, \beta  \in \left[ {0,1} \right]$.
By setting basis ${\left| \varphi \right\rangle _A} = \sin ({\theta
\mathord{\left/
 {\vphantom {\theta  {2)}}} \right.
 \kern-\nulldelimiterspace} {2)}}\left| 0' \right\rangle  + {e^{i\phi }}\cos ({\theta  \mathord{\left/
 {\vphantom {\theta  {2)}}} \right.
 \kern-\nulldelimiterspace} {2)}}\left| 1' \right\rangle ,{\left| {{\varphi ^ \bot }} \right\rangle _A} =  - {e^{ - i\phi }}\cos ({\theta  \mathord{\left/
 {\vphantom {\theta  {2)}}} \right.
 \kern-\nulldelimiterspace} {2)}}\left| 0' \right\rangle  + \sin ({\theta  \mathord{\left/
 {\vphantom {\theta  {2)}}} \right.
 \kern-\nulldelimiterspace} {2)}}\left| 1' \right\rangle $ and considering Eq. (\ref {eq3}), we can get
$LS_A = 2\left| \alpha \right|\left| \beta \right|$ (the detailed
proof of this formula is given in appendix). If $LS_A = 0$ then
$\alpha = 0$ or $\beta = 0$, in either case, ${\left| \psi
\right\rangle _{AB}}$ is a product state. If ${\left| \psi
\right\rangle _{AB}}$ is a product state, then $\alpha = 0$ or
$\beta = 0$, thus $LS_A = 0$. Q.E.D.

\emph{\textbf{Theorem 2:}} For state ${\rho _{AB}}$, the necessary and sufficient condition for $LS_A
= 0$ is that ${\rho _{AB}}$ can be written in the following form
\begin{equation}\label{eq22}
\begin{split}
 {\rho _{AB}} = {p_1}{\left| \varphi \right\rangle _A}\left\langle
\varphi  \right| \otimes \rho _B^1 + {p_2}{\left| {{\varphi ^ \bot
}} \right\rangle _A}\left\langle {{\varphi ^ \bot }} \right| \otimes
\rho _B^2.
\end{split}
\end{equation}

Proof. For mixed state that can be written as the form ${\rho _{AB}} =
{p_1}{\left| \varphi \right\rangle _A}\left\langle \varphi \right|
\otimes \rho _B^1 + {p_2}{\left| {{\varphi ^ \bot }} \right\rangle
_A}\left\langle {{\varphi ^ \bot }} \right| \otimes \rho _B^2\ $,
since basis $\left\{ {{{\left| \varphi  \right\rangle
}_A},\;{{\left| {{\varphi ^ \bot }} \right\rangle }_A}} \right\}$
can be chosen, thus we have $LS_A = 0$.

For states with $LS_A = 0$, from the definition (see Eq.
(\ref{eq3})), we know that there must exist a decomposition ${\rho
_{AB}} = \sum\limits_i {{p_i}{{\left| {{\psi _i}} \right\rangle
}_{AB}}\left\langle {{\psi _i}} \right|} $ and a basis $\left\{
{{{\left| \psi \right\rangle }_A},\;{{\left| {{\psi ^ \bot }}
\right\rangle }_A}} \right\}$ such that for each ${\left| {{\psi
_i}} \right\rangle { _{AB}}}$, $S_A({\left| {{\psi _i}}
\right\rangle _{AB}},\left\{ {{{\left| \varphi  \right\rangle
}_A},{{\left| {{\varphi ^ \bot }} \right\rangle }_A}} \right\}) =
0$. From Eq. (\ref{eq2}), ${\left| {{\psi _i}} \right\rangle _{AB}}$
must can be written as ${\left| {{\psi _i}} \right\rangle _{AB}} =
{\left| {{\psi _i}} \right\rangle _A} \otimes {\left| {{\psi _i}}
\right\rangle _B}$, and ${\left| {{\psi _i}} \right\rangle _A}$ is
either ${{{\left| \varphi \right\rangle }_A}}$ or ${{{\left|
{{\varphi ^ \bot }} \right\rangle }_A}}$. Summing the corresponding
terms with ${\left| \varphi \right\rangle _A}\left\langle \varphi
\right|$ and ${\left| {{\varphi ^ \bot }} \right\rangle
_A}\left\langle {{\varphi ^ \bot }} \right|$, respectively, we have
${\rho _{AB}} = {p_1}{\left| \varphi \right\rangle _A}\left\langle
\varphi \right| \otimes \rho _B^1 + {p_2}{\left| {{\varphi ^ \bot }}
\right\rangle _A}\left\langle {{\varphi ^ \bot }} \right| \otimes
\rho _B^2\ $. Q.E.D.

Note that Eq. (\ref{eq22}) is also the necessary and sufficient
condition for the original definition of quantum discord (when
measurement is done to part A) to be zero \cite{quantum discord},
and it is called a classical-quantum state \cite{QDRMP}.

For any state ${\rho _{AB}}$, the following inequality holds, $0 \le L{S_A}({\rho _{AB}}) \le 1$. The proof is given in appendix.

In the same way, for part B,
\begin{equation}\nonumber
\begin{split}
LS_B({\rho _{AB}}) = \min \sum\limits_i {{p_i}S_B({{\left| {{\psi
_i}} \right\rangle }_{AB}},
\left\{ {{{\left| \varphi  \right\rangle }_B},{{\left| {{\varphi ^ \bot }} \right\rangle }_B}} \right\})}  , \\
\end{split}
\end{equation}
where the minimum is taken over all decompositions $\left\{
{{p_i},\;{{\left| {{\psi _i}} \right\rangle }_{AB}}} \right\}$ and
all basis $\left\{ {{{\left| \varphi  \right\rangle }_B},{{\left|
{{\varphi ^ \bot }} \right\rangle }_B}} \right\}$. The necessary and
sufficient condition for $LS_B = 0$ is that ${\rho _{AB}}$ can be
decomposed in the following form
\begin{equation}\nonumber
\begin{split}
 {\rho _{AB}} = {p_1}\rho _A^1 \otimes {\left| \varphi  \right\rangle _B}\left\langle \varphi  \right| + {p_2}\rho _A^2 \otimes {\left| {{\varphi ^ \bot }} \right\rangle _B}\left\langle {{\varphi ^ \bot }} \right| ,
\end{split}
\end{equation}
which is called a quantum-classical state \cite{QDRMP}. This is also the necessary and sufficient condition for the original definition of quantum discord (when measurement is done to part B) to be zero\cite{quantum discord}.

Note that $LS_A$ and $LS_B$ are neither symmetric. A symmetric one
can be defined
\begin{equation}\label{eq4}
\begin{split}
LS = \min \sum\limits_i {{p_i}}\frac{1}{2} [&S_A({\left| {{\psi _i}}
\right\rangle _{AB}},{\left| \varphi  \right\rangle _A},{\left|
{{\varphi ^ \bot }} \right\rangle _A}) \\
&+ S_B({\left| {{\psi _i}} \right\rangle _{AB}},{\left| \varphi
\right\rangle _B},{\left| {{\varphi ^ \bot }} \right\rangle _B})]
 , \\
\end{split}
\end{equation}
where the minimum is taken over all decompositions $\left\{
{{p_i},\;{{\left| {{\psi _i}} \right\rangle }_{AB}}} \right\}$ and
all basis $\left\{ {{{\left| \psi  \right\rangle }_A},\;{{\left|
{{\psi ^ \bot }} \right\rangle }_A}} \right\}$, $\left\{ {{{\left|
\varphi  \right\rangle }_B},{{\left| {{\varphi ^ \bot }}
\right\rangle }_B}} \right\}$.

\emph{\textbf{Theorem 3: }}For any state ${\rho _{AB}}$, $LS
= 0$ iff it can be written as
\begin{equation}\label{LS0}
{\rho _{AB}} = {p_1}{\left| \varphi  \right\rangle
_A}\left\langle \varphi \right| \otimes {\left| \varphi
\right\rangle _B}\left\langle \varphi \right| + {p_2}{\left|
{{\varphi ^ \bot }} \right\rangle _A}\left\langle {{\varphi ^ \bot
}} \right| \otimes {\left| {{\varphi ^ \bot }} \right\rangle
_B}\left\langle {{\varphi ^ \bot }} \right|.
\end{equation}

The proof is similar to theorem 2, and will be omitted. The state in the right side of Eq. (\ref{LS0}) is a classical state \cite{QDRMP}. This means that for $LS = 0$, the state can be
prepared without quantum operation in either parts, while for $LS
\ne 0$, quantum operation must be introduced to prepare this
state. It is obvious that a state is quantum discordant when and
only when it has local superposition. Note that, generally, $LS
\ge {{(LS_A + LS_B)} \mathord{\left/
 {\vphantom {{(LS_A + LS_B)} 2}} \right.
 \kern-\nulldelimiterspace} 2}$ (the proof will be given in appendix). $LS_A$, $LS_B$, and $LS$ range from $0$ to $1$.

This definition can be easily generalized to multidimensional case.
Consider a two partite system with dimension $d_A$ and $d_B$ for
each subsystem, for state ${\left| {{\psi }} \right\rangle _{AB}}$,
the amount of superposition between basis $\left\{ {{{\left|
{{\varphi ^j}} \right\rangle }_A}} \right\}$ is defined as
\begin{equation}\label{eq20}
\begin{split}
S_A({\left| \psi  \right\rangle _{AB}},\left\{ {{{\left| {{\varphi ^j}} \right\rangle }_A}} \right\}) = 2{\left( {\sum\limits_{m < n} {P(\left| {{\varphi ^m}} \right\rangle )P(\left| {{\varphi ^n}} \right\rangle )} } \right)^{1/2}},  \\
\end{split}
\end{equation}
where $P(\left| {{\varphi ^m}} \right\rangle ) = {\rm Tr}_B\left[
{{}_A{{\left\langle {{{\varphi ^m}}}
 \mathrel{\left | {\vphantom {{{\varphi ^m}} \psi }}
 \right. \kern-\nulldelimiterspace}
 {\psi } \right\rangle }_{AB}}{{\left\langle {\psi }
 \mathrel{\left | {\vphantom {\psi  {{\varphi ^m}}}}
 \right. \kern-\nulldelimiterspace}
 {{{\varphi ^m}}} \right\rangle }_A}} \right]$, $m = 1, \cdots ,{d_A}$. Hence the amount of LS in part A for
 ${\rho _{AB}} = \sum\limits_i {{p_i}{{\left| {{\psi _i}} \right\rangle }_{AB}}\left\langle {{\psi _i}} \right|} $ is
\begin{equation}\label{eq5}
\begin{split}
 LS_A = \min \sum\limits_i {{p_i}S_A({{\left| {{\psi _i}} \right\rangle }_{AB}},\left\{ {{{\left| {{\varphi ^j}} \right\rangle }_A}} \right\})} ,  \\
\end{split}
\end{equation}
where the minimum is taken over all decompositions $\left\{
{{p_i},\;{{\left| {{\psi _i}} \right\rangle }_{AB}}} \right\}$ and
all basis $\left\{ {{{\left| {{\varphi ^j}} \right\rangle }_A}}
\right\}$. In the same way, $LS_B$ for part B and $LS$ for both
parts can be defined analogous to previous bipartite two-state
definitions. Correspondingly, $LS = 0$ holds iff ${\rho _{AB}} =
\sum\limits_{ij} {{p_{ij}}{{\left| {{\varphi ^i}} \right\rangle
}_A}\left\langle {{\varphi ^i}} \right| \otimes {{\left| {{\varphi
^j}} \right\rangle }_B}\left\langle {{\varphi ^j}} \right|}$,
where $\left\{ {{{\left| {{\varphi ^i}} \right\rangle }_A}}
\right\}$ and $\left\{ {{{\left| {{\varphi ^j}} \right\rangle
}_B}} \right\}$ are orthonormal basis for system A and B,
respectively.

Example: For state in $3 \times 3$ system, ${\left| \psi
\right\rangle _{AB}} = \sum\limits_{i,j = 1,2,3} {{a_{ij}}\left|
{ij} \right\rangle } $, by using the parametrization of $3$
dimensional unitary matrix \cite{Un} thus running over all the
basis, the LS in side A or B can be numerically obtained. We compare
the numerical results with $2({\lambda _1}^2{\lambda _2}^2 +
{\lambda _1}^2{\lambda _3}^2 + {\lambda _2}^2{\lambda _3}^2)^{1/2}$,
where $\lambda _1, \lambda _2, \lambda _3$ are the singular value
(also called Schmidt coefficients) of matrix $A$ (with ${a_{ij}}$
being its elements). Specifically, for pure state
\[{\left| \psi  \right\rangle _{AB}} = \sqrt {\frac{2}{3}} \lambda \left| {00} \right\rangle  + \sqrt {\frac{1}{3}} \lambda \left| {11} \right\rangle  + \sqrt {1 - {\lambda ^2}} \left| {22} \right\rangle, \]
where $0 \le \lambda  \le 1$.
\begin{figure}[htbp]
\centering
\includegraphics[width=3.3in]{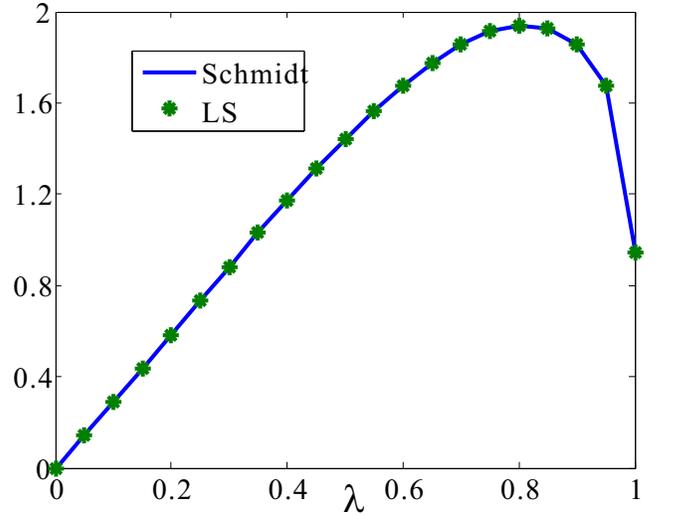}
\caption{\small (color online) For pure state, ${\left| \psi  \right\rangle _{AB}} = \lambda \sqrt 2 /\sqrt 3 \left| {00} \right\rangle  + \lambda 1/\sqrt 3 \left| {11} \right\rangle  + \sqrt {1 - {\lambda ^2}} \left| {22} \right\rangle $, the LS (green star line) compared with $\sqrt 2 /3{\lambda ^2} + 2(\sqrt 2  + 1)\lambda \sqrt {1 - {\lambda ^2}} /\sqrt 3 $(blue solid line)} \label{F-3}
\end{figure}
the value expressed by singular value is $\sqrt 2 /3{\lambda ^2} +
2(\sqrt 2  + 1)\lambda \sqrt {1 - {\lambda ^2}} /\sqrt 3 $. As shown
in Fig. \ref{F-4-3-1} they are the same. For the fifth point in Fig.
\ref{pic1}, $\lambda = 0.2$, $LS = 0.583986531642978$, and results
by Schmidt coefficients equals $0.583986531642978$. However, due to
the complicate calculations, it is hard to strictly prove that $LS =
2({\lambda _1}^2{\lambda _2}^2 + {\lambda _1}^2{\lambda _3}^2 +
{\lambda _2}^2{\lambda _3}^2)^{1/2}$.

\section{Nonlocal superposition in bipartite system}
Consider, for pure state ${\left| \psi  \right\rangle _{AB}} =
{a_{00}}\left| {00} \right\rangle  + {a_{01}}\left| {01}
\right\rangle  + {a_{10}}\left| {10} \right\rangle  +
{a_{11}}\left| {11} \right\rangle $, the state superposition in the product basis, i.e., ${\left| {{\varphi
^{00}}} \right\rangle _{AB}} = {\left| \varphi \right\rangle _A}
\otimes {\left| \varphi \right\rangle _B},\;{\left| {{\varphi
^{10}}} \right\rangle _{AB}} = {\left| {{\varphi ^ \bot }}
\right\rangle _A} \otimes {\left| \varphi \right\rangle
_B},\;{\left| {{\varphi ^{01}}} \right\rangle _{AB}} = {\left|
\varphi  \right\rangle _A} \otimes {\left| {{\varphi ^ \bot }}
\right\rangle _B},\;{\left| {{\varphi ^{11}}} \right\rangle _{AB}}
= {\left| {{\varphi ^ \bot }} \right\rangle _A} \otimes {\left|
{{\varphi ^ \bot }} \right\rangle _B}$. Since all parts of the system are included in the product basis, we define the state
superposition between the product basis as nonlocal superposition.
For state ${\left| \psi \right\rangle _{AB}}$, the amount of
nonlocal superposition in this set of basis is defined as
 \begin{equation}\label{eq30}
\begin{split}
  &NLS({\left| \psi  \right\rangle _{AB}},\left\{ {{{\left| {{\varphi ^{ij}}} \right\rangle }_{AB}}} \right\}) = \\
  &2{\left( {\sum\limits_{2m + n < 2k + l} {P({{\left| {{\varphi ^{mn}}} \right\rangle }_{AB}})P({{\left| {{\varphi ^{kl}}} \right\rangle }_{AB}})} } \right)^{1/2}}, \\
\end{split}
\end{equation}
where $P({\left| {{\varphi ^{mn}}} \right\rangle _{AB}}) =
{}_{AB}{\left\langle {{{\varphi ^{mn}}}}
 \mathrel{\left | {\vphantom {{{\varphi ^{mn}}} \psi }}
 \right. \kern-\nulldelimiterspace}
 {\psi } \right\rangle _{AB}}{\left\langle {\psi }
 \mathrel{\left | {\vphantom {\psi  {{\varphi ^{mn}}}}}
 \right. \kern-\nulldelimiterspace}
 {{{\varphi ^{mn}}}} \right\rangle _{AB}}$, $m,\;n = 0,\;1$, and the sum is taken over all $m, n, k, l$ satisfying ${2m + n < 2k + l}$.

\emph{ \textbf{Definition: }}The amount of NLS of state ${\left| \psi  \right\rangle
 _{AB}}$ is defined as
 \begin{equation}\label{eq6}
\begin{split}
  NLS({\left| \psi  \right\rangle _{AB}}) = \min NLS({\left| \psi  \right\rangle _{AB}},\left\{ {{{\left| {{\varphi ^{ij}}} \right\rangle }_{AB}}} \right\}), \\
\end{split}
\end{equation}
where the minimum is taken over all product basis $\left\{ {{{\left|
{{\varphi ^{ij}}} \right\rangle }_{AB}}} \right\}$.

\emph{\textbf{Theorem 4: }} For pure state, $NLS({\left| \psi \right\rangle
_{AB}}) = 0$ iff ${\left| \psi \right\rangle _{AB}}$ is a product state.

Proof. For product state ${\left| \psi \right\rangle _{AB}}={\left|
\psi \right\rangle _A} \otimes {\left| \psi  \right\rangle _B}$,
product basis $\left\{ {{{\left| {{\varphi ^{ij}}} \right\rangle
}_{AB}}} \right\}$, in which $\left| {{\varphi ^{00}}} \right\rangle
= {\left| \psi \right\rangle _A} \otimes {\left| \psi  \right\rangle
_B}$, can be chosen. Therefore, in this case, there is no state
superposition between the chosen basis, and $NLS({\left| \psi
\right\rangle _{AB}}) = 0$. If $NLS({\left| \psi \right\rangle
_{AB}}) = 0$, from Eq. (\ref{eq6})and Eq. (\ref{30}), there exists a
product basis $\left\{ {{{\left| {{\varphi ^{mn}}} \right\rangle
}_{AB}}} \right\}$ that only one $P({\left| {{\varphi ^{ij}}}
\right\rangle _{AB}})$ is nonzero. Thus, we have ${\left| \psi
\right\rangle _{AB}} = {\left| {{\varphi ^i}} \right\rangle _A}
\otimes {\left| {{\varphi ^j}} \right\rangle _B}$, which is a
product state. Q.E.D.

For $NLS({\left| \psi \right\rangle _{AB}}) \ne 0$,
no matter which set of product basis is chosen, there is state
superposition between the basis.

\begin{figure}[htbp]
\centering
\includegraphics[width=3.3in]{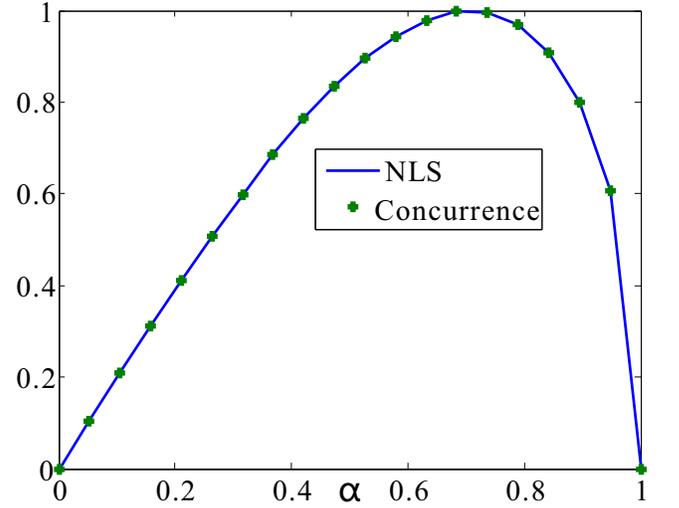}
\caption{\small (color online) $NLS$ (blue solid line) and
Concurrence (asterisk marked) as the function of parameter
$\alpha$.} \label{pic1}
\end{figure}
For any pure state ${\left| \psi  \right\rangle _{AB}} =
{a_{00}}\left| {00} \right\rangle  + {a_{01}}\left| {01}
\right\rangle  + {a_{10}}\left| {10} \right\rangle  + {a_{11}}\left|
{11} \right\rangle$, by Schmidt decomposition, it can be written as
${\left| \psi  \right\rangle _{AB}} = \alpha \left| {0'0'}
\right\rangle  + \beta  \left| {1'1'} \right\rangle  $, where
$\alpha$ and $\beta$ are the singular value of matrix $A$ (with
${a_{ij}}$ being its elements) and $\alpha, \beta  \in \left[ {0,1}
\right]$.  The numeric result of $NLS$ of this pure state is
compared with concurrence \cite{concurrence}, which in this case is
$2 \alpha \beta$. As shown in Fig. \ref{pic1}, $NLS$ and concurrence
are the same. Since for any pure state, $\alpha  \in \left[ {0,1}
\right]$ and $\beta  = \sqrt {1 - {\alpha ^2}}$ holds, the results
in Fig. \ref{pic1} has compared all the pure states. For the fifth
point in Fig. \ref{pic1}, $\alpha = 4/19$, $NLS =
0.411616080243916$, and $Concurrence = 0.411616080243916$. However,
due to the complicate calculations, it is hard to strictly prove
that $NLS({\left| \psi \right\rangle _{AB}}) = 2 \alpha \beta $.

The concurrence, when used
in the definition of entanglement of formation, is only a
mathematical expression. From this state superposition
perspective, concurrence can also be directly considered as the
amount of nonlocal superposition existing in state ${\left| \psi
\right\rangle
 _{AB}}$.

\emph{\textbf{Definition: }}The amount of NLS for mixed state ${\rho
_{AB}}$ is defined as, according to convex roof theory \cite{convex
roof},
 \begin{equation}\label{eq7}
\begin{split}
  NLS({\rho _{AB}}) = \min \sum\limits_i {{p_i}NLS({{\left| {{\psi _i}} \right\rangle }_{AB}})} , \\
\end{split}
\end{equation}
where the minimum is taken over all decompositions $\left\{
{{p_i},\;{{\left| {{\psi _i}} \right\rangle }_{AB}}} \right\}$.

Note
that the minimum taken here is different from that in LS. In NLS,
local quantum operation, which will not introduce any nonlocal
superposition, should be allowed. By local unitary operation, any
two different product basis can be changed to each other, e.g., for
$\left\{ {{{\left| {{\varphi ^{ij}}} \right\rangle }_{AB}} =
{{\left| {{\varphi ^i}} \right\rangle }_A} \otimes {{\left|
{{\varphi ^j}} \right\rangle }_B}\;} \right\}$ and $\left\{
{{{\left| {{\varphi ^{'ij}}} \right\rangle }_{AB}} = {{\left|
{{\varphi ^{'i}}} \right\rangle }_A} \otimes {{\left| {{\varphi
^{'j}}} \right\rangle }_B}} \right\}$, by local unitary operation in
part A, we can change $\left\{ {{{\left| {{\varphi ^i}}
\right\rangle }_A}\;} \right\}$ to $\left\{ {{{\left| {{\varphi
^{'i}}} \right\rangle }_A}} \right\}$, and the same way for part B.
Therefore, in NLS (Eq. (\ref{eq7})) for each ${{{\left| {{\psi _i}}
\right\rangle }_{AB}}}$, unlike the minimum in the definition of LS
(Eq. (\ref{eq3})) where the basis is the same for every ${{{\left|
{{\psi _i}} \right\rangle }_{AB}}}$, different product basis should
be allowed.

From the perspective of preparation, as showed above, by local quantum operation, we have the choice to choose product basis (without introducing any nonlocal superposition) for the preparation of each pure state ensemble ${\left| {{\psi _i}} \right\rangle _{AB}}$. Thus, Eq. (\ref{eq7}) gives the minimum nonlocal superposition produced in the preparation. When $NLS({\rho _{AB}}) = 0$, by choosing certain decomposition and basis, local quantum operation is enough to prepare the state, while when $NLS({\rho _{AB}}) \ne 0$, no matter which decomposition and product basis are chosen, nonlocal quantum operation must be introduced.

\emph{\textbf{Theorem 5:}} For any state ${\rho _{AB}}$, $NLS({\rho _{AB}}) = 0$ iff $\rho _{AB}$
is separable ($\rho _{AB}$ can be written as ${\rho _{AB}} = \sum\limits_k {{p_k}\rho _A^k \otimes \rho _B^k} $).

Proof. If ${\rho _{AB}} = \sum\limits_k {{p_k}\rho _A^k \otimes \rho
_B^k} $, each ${\rho _A^k}$ and  ${\rho _B^k}$ can be decomposed
again into pure state ensemble, $\rho _A^k = \sum\limits_i
{p_{kA}^i{{\left| {{\varphi ^{ki}}} \right\rangle }_A}\left\langle
{{\varphi ^{ki}}} \right|} $, $\rho _B^k = \sum\limits_i
{p_{kB}^i{{\left| {{\varphi ^{ki}}} \right\rangle }_B}\left\langle
{{\varphi ^{ki}}} \right|} $, where $\left\{ {{{\left| {{\varphi
^{ki}}} \right\rangle }_A}} \right\}$ and $\left\{ {{{\left|
{{\varphi ^{ki}}} \right\rangle }_B}} \right\}$ are orthogonal basis
for each system. Thus, ${\rho _{AB}}$ can be decomposed as ${\rho
_{AB}} = \sum\limits_{k,i,j} {{p_k}p_{kA}^ip_{kA}^i{{\left|
{{\varphi ^{ki}}} \right\rangle }_A}\left\langle {{\varphi ^{ki}}}
\right| \otimes {{\left| {{\varphi ^{ki}}} \right\rangle
}_B}\left\langle {{\varphi ^{ki}}} \right|} $. Under this
decomposition, each pure state ensemble is a pure product state,
which, according to theorem 4, has zero nonlocal superposition.
Thus, if $\rho _{AB}$ is separable, $NLS({\rho _{AB}}) = 0$.

If $NLS({\rho _{AB}}) = 0$, according to Eq. (\ref{eq7}), there exists a certain decomposition $\left\{ {{p_i},{{\left| {{\psi _i}} \right\rangle }_{AB}}} \right\}$, that for each ${{{\left| {{\psi _i}} \right\rangle }_{AB}}}$, $NLS({\left| {{\psi _i}} \right\rangle _{AB}}) = 0$. According to theorem 4, each ${{{\left| {{\psi _i}} \right\rangle }_{AB}}}$ is a product state, ${\left| {{\psi _i}} \right\rangle _{AB}} = {\left| {{\varphi _i}} \right\rangle _A} \otimes {\left| {{\varphi _i}} \right\rangle _B}$. Thus, $\rho _{AB}$ can be written as ${\rho _{AB}} = \sum\limits_i {{p_i}{{\left| {{\varphi _i}} \right\rangle }_A}\left\langle {{\varphi _i}} \right| \otimes {{\left| {{\varphi _i}} \right\rangle }_B}\left\langle {{\varphi _i}} \right|} $, which is a separable state. Q.E.D.

Theorem 5 directly indicates that a state is entangled when and only
when it has nonlocal superposition. Thus, from this superposition perspective, entanglement can be considered as the nonlocal superposition existing in the system. Besides, as we numerically showed above, for pure state, concurrence and NLS are the same. According to the way concurrence calculated for mixed states in Ref. \cite{concurrence} , for any two qubits,
NLS has the same value as concurrence.

This definition can be easily generalized to multidimensional
states. Consider a bipartite system  with dimension $d_A$ and
$d_B$ for each subsystem, for state ${\left| \psi  \right\rangle
_{AB}} = \sum\limits_{i = 1}^{{d_A}} {\sum\limits_{j = 1}^{{d_B}}
{{a_{ij}}{{\left| i \right\rangle }_A}{{\left| j \right\rangle }_B}}
} $, the amount of NLS between product basis ${\left| {{\varphi
^{mn}}} \right\rangle _{AB}} = {\left| {{\varphi ^m}} \right\rangle
_A} \otimes {\left| {{\varphi ^n}} \right\rangle _B}\;$, where $m =
1, \cdots ,{d_A}$, $n = 1, \cdots ,{d_B}$ and $\left\{ {{{\left|
{{\varphi ^m}} \right\rangle }_A}} \right\}$, $\left\{ {{{\left|
{{\varphi ^n}} \right\rangle }_B}} \right\}$ are othonormal basis of
system A and B, respectively, is defined as
 \begin{equation}\nonumber
\begin{split}
&NLS({\left| \psi  \right\rangle _{AB}},\left\{ {{{\left| {{\varphi ^{ij}}} \right\rangle }_{AB}}} \right\}) = \\
&2{\left( {\sum\limits_{m{d_B} + n < k{d_B} + l} {P({{\left| {{\varphi ^{mn}}} \right\rangle }_{AB}})P({{\left| {{\varphi ^{kl}}} \right\rangle }_{AB}})} } \right)^{1/2}},\\
\end{split}
\end{equation}
where $P({\left| {{\varphi ^{mn}}} \right\rangle _{AB}}) =
{}_{AB}{\left\langle {{{\varphi ^{mn}}}}
 \mathrel{\left | {\vphantom {{{\varphi ^{mn}}} \psi }}
 \right. \kern-\nulldelimiterspace}
 {\psi } \right\rangle _{AB}}{\left\langle {\psi }
 \mathrel{\left | {\vphantom {\psi  {{\varphi ^{mn}}}}}
 \right. \kern-\nulldelimiterspace}
 {{{\varphi ^{mn}}}} \right\rangle _{AB}}$ and the summation is taken over all $m, n, k, l$ satisfying $md_B + n < kd_B + l$. The amount of NLS of state ${\left| \psi  \right\rangle _{AB}}$ is
\begin{equation}\label{eq8}
\begin{split}
 NLS({\left| \psi  \right\rangle _{AB}}) = \min
NLS(\left\{ {{{\left| {{\varphi ^{ij}}} \right\rangle }_{AB}}}
\right\}) , \\
\end{split}
\end{equation}
where the minimum is taken over all product basis $\left\{ {{{\left|
{{\varphi ^{ij}}} \right\rangle }_{AB}}} \right\}$. Using convex
roof theory, NLS for mixed states can also be defined.

\section{Local and nonlocal superposition in multipartite system}
The definition of
LS and NLS can be generalized to multipartite case. Assuming an $n$
partite system with dimension $d_m$ for the $m$th subsystem, for
pure state
\begin{equation}\nonumber
\begin{split}
{\left| \psi  \right\rangle _{{i_1}, \ldots ,{i_n}}} =
\sum\limits_{{i_1}, \ldots ,{i_n}} {{a_{{i_1}, \ldots ,{i_n}}}
{{\left| i \right\rangle }_1} \otimes  \ldots  \otimes {{\left| i \right\rangle }_n}} , \\
\end{split}
\end{equation}
where ${i_m} = 1, \ldots ,{d_m}$, the amount of LS for the $m$th
part, when basis $\left\{ {{{\left| {{\varphi ^{{i_m}}}}
\right\rangle }_m}} \right\}$ is chosen, reads
\begin{equation}\nonumber
\begin{split}
&LS_m({\left| \psi  \right\rangle _{1, \ldots ,n}},\left\{ {{{\left| {{\varphi ^{{i_m}}}} \right\rangle }_m}} \right\}) = \\
&2{\left( {\sum\limits_{{k_m} < {l_m}} {P({{\left| {{\varphi ^{{k_m}}}} \right\rangle }_m})P({{\left| {{\varphi ^{{l_m}}}} \right\rangle }_m})} } \right)^{1/2}}, \\
\end{split}
\end{equation}
where $P({\left| {{\varphi ^{{k_m}}}} \right\rangle _m}) = {\rm
Tr}_{1, \ldots ,m - 1,m + 1, \ldots ,n}[{}_m{\left\langle {{{\varphi
^{{k_m}}}}}
 \mathrel{\left | {\vphantom {{{\varphi ^{{k_m}}}} \psi }}
 \right. \kern-\nulldelimiterspace}
 {\psi } \right\rangle _{1, \ldots ,n}}\\
 {\left\langle {\psi }
 \mathrel{\left | {\vphantom {\psi  {{\varphi ^{{k_m}}}}}}
 \right. \kern-\nulldelimiterspace}
 {{{\varphi ^{{k_m}}}}} \right\rangle _m}]$. For a mixed state ${\rho _{1, \ldots ,n}} = \sum\limits_i {{p_i}{{\left| {{\psi _i}} \right\rangle }_{1, \ldots ,n}}\left\langle {{\psi _i}} \right|} $, the amount of LS
 for the
$m$th part is defined as
\begin{equation}\label{eq9}
\begin{split}
LS_m({\rho _{1, \ldots ,n}}) = \min \sum\limits_i {{p_i}LS_m({{\left| {{\psi _i}} \right\rangle }_{1, \ldots ,n}},\left\{ {{{\left| {{\varphi ^{{i_m}}}} \right\rangle }_m}} \right\})}   , \\
\end{split}
\end{equation}
where the minimum is taken over all decompositions $\left\{
{{p_i},\;{{\left| {{\psi _i}} \right\rangle }_{1, \ldots ,n}}}
\right\}$ and all basis $\left\{ {{{\left| {{\psi ^{{j_m}}}}
\right\rangle }_m}} \right\}$. The amount of LS including all parts
is
\begin{equation}\label{eq10}
\begin{split}
LS = \min \sum\limits_i {{p_i}\frac{1}{n}\sum\limits_m {LS_m({{\left| {{\psi _i}} \right\rangle }_{1, \ldots ,n}},\left\{ {{{\left| {{\varphi ^{{i_m}}}} \right\rangle }_m}} \right\})} }  , \\
\end{split}
\end{equation}
where the minimum is taken over all decompositions $\left\{
{{p_i},\;{{\left| {{\psi _i}} \right\rangle }_{1, \ldots ,n}}}
\right\}$ and all basis $\left\{ {{{\left| {{\psi ^{{j_1}}}}
\right\rangle }_1}} \right\}$, . . . ,$\left\{ {{{\left| {{\psi
^{{j_n}}}} \right\rangle }_n}} \right\}$. Eq. (\ref{eq10})
indicates that $LS = 0$ iff ${\rho _{1, \ldots ,n}}$ can be
written as ${\rho _{1, \ldots ,n}} = \sum\limits_{{i_1}_1, \ldots
,{i_n}} {{p_{{i_1}, \ldots ,{i_n}}}\left| {{i_1}} \right\rangle
\left\langle {{i_1}} \right| \otimes  \cdots  \otimes \left|
{{i_n}} \right\rangle \left\langle {{i_n}} \right|} $, where
$\left\{ {\left| {{i_m}} \right\rangle } \right\}$ is the
orthonormal basis of part $m$.

For NLS, for pure state ${\left| \psi \right\rangle _{{1}, \ldots
,{n}}}$, when choosing product basis $\left| {{\varphi ^{{i_1},
\ldots ,{i_n}}}} \right\rangle  = \left| {\varphi _1^{{i_1}}}
\right\rangle \otimes \ldots  \otimes \left| {\varphi _n^{{i_n}}}
\right\rangle $, where $\left\{ {{{\left| {{\varphi ^{{i_m}}}}
\right\rangle }_m}} \right\}$ are the orthonormal basis for part
$m$, the amount of NLS under this product basis is defined as
\begin{equation}\nonumber
\begin{split}
&NLS({\left| \psi  \right\rangle _{1, \ldots ,n}},\left\{ {\left| {{\varphi ^{1, \ldots ,n}}} \right\rangle } \right\}) =\\
 &2{\left( {\sum\limits_{R < {R^{'}}} {P(\left| {{\varphi ^{{i_1}, \ldots ,{i_n}}}} \right\rangle )P(\left| {{\varphi ^{i_1^{'}, \ldots ,i_n^{'}}}} \right\rangle )} } \right)^{1/2}}, \\
\end{split}
\end{equation}
where $R = \sum\limits_k {({i_k} - 1) * {d_{k + 1}} *  \cdots  *
{d_n}} $ and $P(\left| {{\varphi ^{{i_1}, \ldots ,{i_n}}}}
\right\rangle ) = {\left\langle {{{\varphi ^{{i_1}, \ldots
,{i_n}}}}}
 \mathrel{\left | {\vphantom {{{\varphi ^{{i_1}, \ldots ,{i_n}}}} \psi }}
 \right. \kern-\nulldelimiterspace}
 {\psi } \right\rangle _{1, \ldots ,n}}\left\langle {\psi }
 \mathrel{\left | {\vphantom {\psi  {{\varphi ^{{i_1}, \ldots ,{i_n}}}}}}
 \right. \kern-\nulldelimiterspace}
 {{{\varphi ^{{i_1}, \ldots ,{i_n}}}}} \right\rangle $. The amount of NLS for state ${\left| \psi \right\rangle
_{{i_1}, \ldots ,{i_n}}}$ is
\begin{equation}\label{eq11}
\begin{split}
NLS({\left| \psi  \right\rangle _{1, \ldots ,n}}) = \min NLS({\left| \psi  \right\rangle _{1, \ldots ,n}},\left\{ {\left| {{\varphi ^{{i_1}, \ldots ,{i_n}}}} \right\rangle } \right\}) , \\
\end{split}
\end{equation}
where the minimum is taken over all product basis $\left\{ {\left|
{{\varphi ^{{i_1}, \ldots ,{i_n}}}} \right\rangle } \right\}$.

For mixed state ${\rho _{1, \ldots ,n}}$, the amount of NLS is
\begin{equation}\label{eq12}
\begin{split}
NLS({\rho _{1, \ldots ,n}}) = \min \sum\limits_i {{p_i}NLS({{\left| {{\psi _i}} \right\rangle }_{1, \ldots ,n}})} , \\
\end{split}
\end{equation}
where the minimum is taken over all decompositions $\left\{
{{p_i},\;{{\left| {{\psi _i}} \right\rangle }_{1, \ldots ,n}}}
\right\}$. Eq. (\ref{eq12}) indicates that $NLS({\rho _{1, \ldots
,n}}) = 0$ iff ${\rho _{1, \ldots ,n}} = \sum\limits_i {{p_i}\rho
_1^i \otimes \cdots \otimes \rho _n^i} $, which is fullly separable. The proof is similar to theorem 5.

LS and NLS can also be defined in partial separable ways. If the
Hilbert space is divided as $\left\{ {{I_1}, \ldots ,{I_k}}
\right\}$, where ${I_i}$ is independent subset of $I = \left\{ {1,
\ldots ,n} \right\}$ and $ \cup _{l = 1}^k{I_l} = I$, LS and NLS can
be defined in the above ways by changing the system to this $k$
partite.  $NLS({\rho _{I_1, \ldots ,I_k}}) = 0$ if and only if the
state is $k$ partite separable respect to the above partition.

As defined above in this section, the superpositions existing in
multipartite system are plural. It can be in a single part, in some
parts, or in all parts of the system. Table 1 shows all the
superpositions in a three partite system. In the table, $|$
represents the partition of the system, e.g., AB$|$C represents
taking part A and part B as a whole and dividing the system as part
AB and part C. For the examples, ${\left| GHZ  \right\rangle } =
\frac{1}{{\sqrt 2 }}{\left| 0 \right\rangle _A}{\left| 0
\right\rangle _B}{\left| 0 \right\rangle _C} + \frac{1}{{\sqrt 2
}}{\left| 1 \right\rangle _A}{\left| 1 \right\rangle _B}{\left| 1
\right\rangle _C}$, ${\left| W  \right\rangle} = \frac{1}{{\sqrt 3
}}{\left| 1 \right\rangle _A}{\left| 0 \right\rangle _B}{\left| 0
\right\rangle _C} + \frac{1}{{\sqrt 3 }}{\left| 0 \right\rangle
_A}{\left| 1 \right\rangle _B}{\left| 0 \right\rangle _C} +
\frac{1}{{\sqrt 3 }}{\left| 0 \right\rangle _A}{\left| 0
\right\rangle _B}{\left| 1 \right\rangle _C}$. These can be easily
generalized to $n (n > 3)$ partite system.

\begin{table}[tbp]
\renewcommand\arraystretch{1.5}
\centering
\begin{tabular}{lccc}
\hline
Superposition &Basis &$\left| GHZ \right\rangle$ &$\left| W \right\rangle$\\
\hline
NLS in A$|$B$|$C &$\left\{ {{{\left| \varphi  \right\rangle }_A} \otimes {{\left| \varphi  \right\rangle }_B} \otimes {{\left| \varphi  \right\rangle }_C}} \right\}$ &1 &1.155\\
NLS in AB$|$C &$\left\{ {{{\left| \varphi  \right\rangle }_{AB}} \otimes {{\left| \varphi  \right\rangle }_C}} \right\}$ &1 &0.943\\
NLS in AB$|$C &$\left\{ {{{\left| \varphi  \right\rangle }_A} \otimes {{\left| \varphi  \right\rangle }_{BC}}} \right\}$ &1 &0.943\\
NLS in AC$|$B &$\left\{ {{{\left| \varphi  \right\rangle }_{AC}} \otimes {{\left| \varphi  \right\rangle }_B}} \right\}$ &1 &0.943\\
LS in A &$\left\{ {{{\left| \varphi  \right\rangle }_A}} \right\}$ &1 &0.943\\
LS in B &$\left\{ {{{\left| \varphi  \right\rangle }_B}} \right\}$ &1 &0.943\\
LS in C &$\left\{ {{{\left| \varphi  \right\rangle }_C}} \right\}$ &1 &0.943\\
LS in AB &$\left\{ {{{\left| \varphi  \right\rangle }_{AB}}} \right\}$ &1 &0.943\\
LS in AC &$\left\{ {{{\left| \varphi  \right\rangle }_{AC}}} \right\}$ &1 &0.943\\
LS in BC &$\left\{ {{{\left| \varphi  \right\rangle }_{BC}}} \right\}$ &1 &0.943\\
\hline
\end{tabular}
\caption{Superpositions existing in three partite system. ${\left| GHZ  \right\rangle } = \frac{1}{{\sqrt 2 }}{\left| 0 \right\rangle _A}{\left| 0 \right\rangle _B}{\left| 0 \right\rangle _C} + \frac{1}{{\sqrt 2 }}{\left| 1 \right\rangle _A}{\left| 1 \right\rangle _B}{\left| 1 \right\rangle _C}$, ${\left| W  \right\rangle} = \frac{1}{{\sqrt 3 }}{\left| 1 \right\rangle _A}{\left| 0 \right\rangle _B}{\left| 0 \right\rangle _C} + \frac{1}{{\sqrt 3 }}{\left| 0 \right\rangle _A}{\left| 1 \right\rangle _B}{\left| 0 \right\rangle _C} + \frac{1}{{\sqrt 3 }}{\left| 0 \right\rangle _A}{\left| 0 \right\rangle _B}{\left| 1 \right\rangle _C}$.}
\end{table}

\begin{figure}[htbp]
\centering
\includegraphics[width=3.3in]{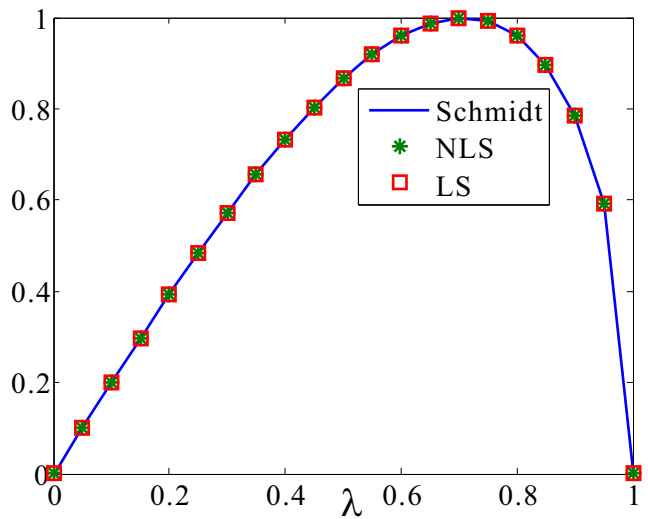}
\caption{\small (color online) GHZ-like states. $NLS$ (asterisk
marked), $LS$ (square marked) and  $2\lambda \sqrt {1 - {\lambda
^2}} $ (blue solid) as the function of parameter $\lambda$.}
\label{pic2}
\end{figure}

Examples: For GHZ-like states $\left| \psi \right\rangle  =
\lambda {\left| 0 \right\rangle _A}{\left| 0 \right\rangle
_B}{\left| 0 \right\rangle _C} + \sqrt {1 - {\lambda ^2}} {\left|
1 \right\rangle _A}{\left| 1 \right\rangle _B}{\left| 1
\right\rangle _C}$. This state has Schmidt decomposition and is already written in that form. The $LS$ and $NLS$ for this state is shown in
Fig. \ref{pic2} compared with the expression calculated by Schmidt
value $2\lambda \sqrt {1 - {\lambda ^2}} $.

\begin{figure}[htbp] \centering
\includegraphics[width=3.3in]{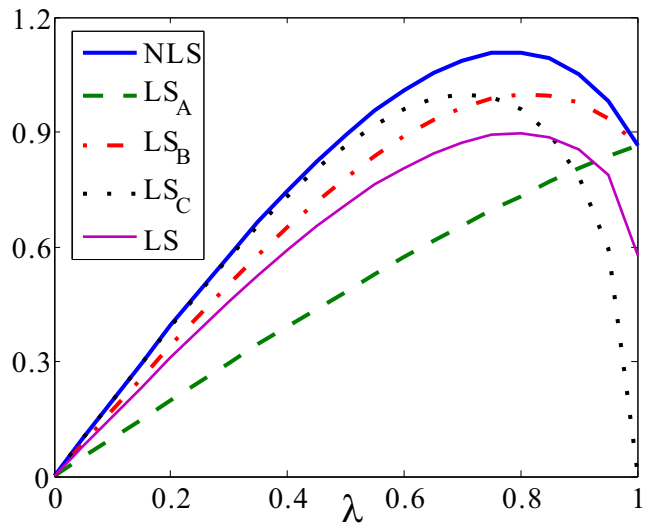}
\caption{\small (color online) W-like states. $NLS$ (blue bold soLS
line), $LS_A$ (green dashed line), $LS_B$ (red dash-dotted line),
$LS_C$ (dotted black line), and $LS$ (purple thin solid line) as the
function of parameter $\lambda$.} \label{pic3}
\end{figure}

For W-like states
$\left| \psi  \right\rangle  = {\lambda  \mathord{\left/
 {\vphantom {\lambda  2}} \right.
 \kern-\nulldelimiterspace} 2}{\left| 0 \right\rangle _A}{\left| 0 \right\rangle _B}{\left| 1 \right\rangle _C} + {{\sqrt 3 \lambda } \mathord{\left/
 {\vphantom {{\sqrt 3 \lambda } 2}} \right.
 \kern-\nulldelimiterspace} 2}{\left| 0 \right\rangle _A}{\left| 1 \right\rangle _B}{\left| 0 \right\rangle _C}
 + \sqrt {1 - {\lambda ^2}} {\left| 1 \right\rangle _A}{\left| 0 \right\rangle _B}{\left| 0 \right\rangle _C}$. Unlike the two partite states and GHZ-like
states, there is no Schmidt decomposition of this state. The
$LS_A$, $LS_B$, $LS_C$, $LS$, and $NLS$ is shown in Fig.
\ref{pic3}. It can be seen that in three partite case, when there
is no Schmidt decomposition, the $LS$ and $NLS$ might be different.

\section{Discussion}
The distinctive feature of quantum world different from classical
world is state superposition. For composite system, the state
superposition can exist between basis of a single part or product
basis including two or more parts. These superpositions are the quantum resources used in quantum information processing. Quantum entanglement and quantum discord have both been considered as quantum resources. As we have showed in this paper, state with local superposition is equivalent to state nonclassical, and state with nonlocal superposition is equivalent to state entangled. From this superposition perspective, the quantum entanglement resource can be considered as the nonlocal superposition in system, while for quantum discord resource, it includes the superposition in single part. However, whether the several kinds of discord introduced before only includes the local superposition is still an open question. From this state superposition aspect, quantum resources are categorized by the superpositions in different parts, and quantum entanglement and quantum discord both capture some kind of specific superposition in the system.

For pure states, we showed that when Schmidt decomposition exists,
the amount of LS is the same as NLS. Thus, for these states, the
quantification of entanglement can be reduced to the property of the
reduced state of a single part, e.g. the entanglement of formation
for two partite pure state is defined by the reduced density matrix
of either part \cite{EOF}. However, as we showed, for pure states
without Schmidt decomposition, the amount of LS and NLS might be
different. For these states, it is inconvenient to study
entanglement by the reduced density matrix, and the state
superposition view introduced in this paper might be useful.

It should be noted that, in Eq. (\ref{eq20}) and Eq. (\ref{eq30}),
although one kind of measure of superposition is given in the
expressions, the specific mathematical formula can be changed. For
the mathematical form defined in this paper, we have not yet find
the way to calculate the local superposition for mixed two partite
states due to the difficulty of taking over all decompositions.
Whether there is a way to calculate the local superposition with the
mathematical form presented in this paper or there are some other
reasonable mathematical definitions that can make all the local
superposition and nonlocal superposition easier to calculate is
still an open question.

The applications of quantum resources in quantum information processes can also be seen from this state superposition aspect. As state entangled is equivalent to state with nonzero NLS, quantum process, for which quantum entanglement is necessary, such as entanglement swapping \cite{ES1}, can also be considered as having explored the nonlocal superpositions in the system.
And the states with zero
NLS but nonzero LS can also be useful in some quantum process. For example, the state
$\rho = \frac{1}{2}\left| 0 \right\rangle \left\langle 0 \right|
\otimes \left|  - \right\rangle \left\langle -  \right| +
\frac{1}{2}\left| + \right\rangle \left\langle  + \right| \otimes
\left| 1 \right\rangle \left\langle 1 \right|$, with no nonlocal
state superposition but nonzero local state superposition (the proof that this state has nonzero LS is presented in appendix), can be used for remote state preparation, as shown in Ref. \cite{RSPNP}. For multipartite case, the
superpositions might be very rich, which might exist in any single
part, any two partite, or any combination of them. These plural
superpositions provide us tremendous ways to explore the quantum
feature of quantum systems. Thus, from the perspective of state superposition,
our results are useful for the consideration of resource
for quantum process.

\section{Conclusion}
We have studied quantum resources from the perspective of quantum state superposition. We have given clear definition and quantification of local superposition and nonlocal superposition.
For states in composite
system, the LS is defined as the superposition between basis of a single
part and NLS as the superposition between product basis of all
parts. From the quantum state superposition perspective, quantum resources are categorized by superpositions existing in different parts. From the aspect of preparation, when nonzero LS presented, quantum operation must be introduced to prepare this state, and when nonzero NLS presented, nonlocal quantum operation must be introduced. We showed that state with zero local superposition is equivalent to state classical and state with zero nonlocal superposition is equivalent to state separable. From this aspect, the quantum entanglement resource can be considered as the nonlocal superposition in system, while for quantum discord resource, we only know that it includes the superposition in single part, and whether it also includes some nonlocal superposition is an open question. From this aspect, the difference between quantum entanglement and quantum discord appears clear.

Besides, LS and NLS are defined in multipartite case. For this case, the kinds of superpositions are plural. We
show that for three partite pure state, when there is no Schmidt
decomposition, the amount of LS and NLS might be different and this
state superposition view might be useful in studying multipartite
entanglement. All these results provide us a direction for the
consideration of states proper for specific quantum
information processes.

This work is supported by the National Science Fund for
Distinguished Young Scholars of China (Grant No. 61225003), National
Natural Science Foundation of China (Grant No. 61101081), and the
National Hi-Tech Research and Development (863) Program.

\begin{appendix}
\section*{Appendix}

1. Detailed proof for the equality $LS_A = 2\left| \alpha \right|\left| \beta \right|$ presented in the proof of theorem 1.

First, write the state as ${\left| \psi  \right\rangle _{AB}} = \alpha \left| {0'0'} \right\rangle  + \beta \left| {1'1'} \right\rangle $. Since this is a pure state, there is only one kind of decomposition ${\left| \psi  \right\rangle _{AB}}\left\langle \psi  \right|$. Thus, to get the minimum in Eq. (\ref{eq3}), we only need to take over all the basis.  Consider the following form of basis:
\[\begin{array}{l}
{\left| \varphi  \right\rangle _A} = \sin \frac{\theta }{2}\left| {0'} \right\rangle  + \cos \frac{\theta }{2}{e^{i\phi }}\left| {1'} \right\rangle ,\\
{\left| {{\varphi ^ \bot }} \right\rangle _A} =  - \cos \frac{\theta }{2}{e^{ - i\phi }}\left| {0'} \right\rangle  + \sin \frac{\theta }{2}\left| {1'} \right\rangle,
\end{array}\]
where $0 \le \theta  \le \pi ,0 \le \phi  \le 2\pi $. By taking all the values of $\theta ,\phi $, all the basis are taken.
According to the definition of $P({\left| \varphi  \right\rangle _A})$ and $P({\left| {{\varphi ^ \bot }} \right\rangle _A})$ in Eq. (\ref{eq2}) , we have
 \[\begin{array}{l}
P({\left| \varphi  \right\rangle _A}) = {\left| \alpha  \right|^2}{\sin ^2}\frac{\theta }{2} + {\left| \beta  \right|^2}{\cos ^2}\frac{\theta }{2}\\
P({\left| {{\varphi ^ \bot }} \right\rangle _A}) = {\left| \alpha  \right|^2}{\cos ^2}\frac{\theta }{2} + {\left| \beta  \right|^2}{\sin ^2}\frac{\theta }{2}.
\end{array}\]
By multiplying them,
\begin{equation}\nonumber
\begin{split}
P({\left| \varphi  \right\rangle _A})P({\left| {{\varphi ^ \bot }} \right\rangle _A}) = &({\left| \alpha  \right|^4} + {\left| \beta  \right|^4}){\sin ^2}\frac{\theta }{2}{\cos ^2}\frac{\theta }{2} \\
&+ {\left| \alpha  \right|^2}{\left| \beta  \right|^2}({\sin ^4}\frac{\theta }{2} + {\cos ^4}\frac{\theta }{2}).
\end{split}
\end{equation}
Considering that ${\cos ^2}\frac{\theta }{2} = 1 - {\sin ^2}\frac{\theta }{2}$,
\begin{equation}\nonumber
\begin{split}
P({\left| \varphi  \right\rangle _A})P({\left| {{\varphi ^ \bot }} \right\rangle _A}) = & - {({\left| \alpha  \right|^2} - {\left| \beta  \right|^2})^2}\left[ {{{({{\sin }^2}\frac{\theta }{2} - \frac{1}{2})}^2} + \frac{1}{4}} \right] \\
&+ {\left| \alpha  \right|^2}{\left| \beta  \right|^2}.
\end{split}
\end{equation}
The above formula get the minimum when
${\sin ^2}\frac{\theta }{2} $ equals $1$ or $0$ (with basis $\left\{ {\left| {0'} \right\rangle ,\left| {1'} \right\rangle } \right\}$), and the minimum is ${\left| \alpha  \right|^2}{\left| \beta  \right|^2}$.
Thus, according to Eq. (\ref{eq3}), we have $LS_A = 2\left| \alpha \right|\left| \beta \right|$. Q.E.D.

2. Proof for $0 \le L{S_A}({\rho _{AB}}) \le 1$.

 For pure state, we have proved before that $L{S_A}({\left| \psi  \right\rangle _{AB}}) = 2\left| \alpha  \right|\left| \beta  \right|$. Since $0 \le 2\left| \alpha  \right|\left| \beta  \right| \le 1$, thus, for pure state, $0 \le L{S_A}({\left| \psi  \right\rangle _{AB}}) \le 1$.

 For mixed state ${\rho _{AB}}$, we first prove that for ${\left| \psi  \right\rangle _{AB}}$, the local superposition between any basis is less than or equal $1$. For any basis $\left\{ {{{\left| \varphi  \right\rangle }_A},{{\left| {{\varphi ^ \bot }} \right\rangle }_A}} \right\}$, we have
 \[2\sqrt {P({{\left| \varphi  \right\rangle }_A})P({{\left| {{\varphi ^ \bot }} \right\rangle }_A})}  \le \left( {P({{\left| \varphi  \right\rangle }_A}) + P({{\left| {{\varphi ^ \bot }} \right\rangle }_A})} \right) = 1.\]
Thus, ${S_A}(\left| \psi  \right\rangle _{AB},\left\{ {{{\left| \varphi  \right\rangle }_A},{{\left| {{\varphi ^ \bot }} \right\rangle }_A}} \right\}) \le 1$. For any decomposition $\left\{ {{p_i},\;{{\left| {{\psi _i}} \right\rangle }_{AB}}} \right\}$,
 \[\sum\limits_i {{p_i}{S_A}({{\left| {{\psi _i}} \right\rangle }_{AB}},\left\{ {{{\left| \varphi  \right\rangle }_A},{{\left| {{\varphi ^ \bot }} \right\rangle }_A}} \right\})}  \le \sum\limits_i {{p_i}}  = 1.\]
Thus,
 \[\min \sum\limits_i {{p_i}{S_A}({{\left| {{\psi _i}} \right\rangle }_{AB}},\left\{ {{{\left| \varphi  \right\rangle }_A},{{\left| {{\varphi ^ \bot }} \right\rangle }_A}} \right\})}  \le 1.\]
Q.E.D.

3. Proof for $LS
\ge {{(LS_A + LS_B)} \mathord{\left/
 {\vphantom {{(LS_A + LS_B)} 2}} \right.
 \kern-\nulldelimiterspace} 2}$.

 According to the definition of local superposition in Eq.  (\ref{eq4}),
\begin{equation}\nonumber
\begin{split}
LS = &\min\frac{1}{2} \sum\limits_i{{p_i}} [{S_A}({\left| {{\psi _i}} \right\rangle _{AB}},{\left| \varphi  \right\rangle _A},{\left| {{\varphi ^ \bot }} \right\rangle _A})\\
            &+ {S_B}({\left| {{\psi _i}} \right\rangle _{AB}},{\left| \varphi  \right\rangle _B},{\left| {{\varphi ^ \bot }} \right\rangle _B})], \\
\end{split}
\end{equation}
where the minimum is taken over all decompositions $\left\{
{{p_i},\;{{\left| {{\psi _i}} \right\rangle }_{AB}}} \right\}$ and
all basis $\left\{ {{{\left| \psi  \right\rangle }_A},\;{{\left|
{{\psi ^ \bot }} \right\rangle }_A}} \right\}$, $\left\{ {{{\left|
\varphi  \right\rangle }_B},{{\left| {{\varphi ^ \bot }}
\right\rangle }_B}} \right\}$. It is straight forward that
\begin{equation}\nonumber
\begin{split}
 LS \ge &\frac{1}{2}\min \sum\limits_i {{p_i}} {S_A}({\left| {{\psi _i}} \right\rangle _{AB}},{\left| \varphi  \right\rangle _A},{\left| {{\varphi ^ \bot }} \right\rangle _A}) \\
 &+ \frac{1}{2}\min \sum\limits_i {p_i^{'}} {S_B}({\left| {\psi _i^{'}} \right\rangle _{AB}},{\left| \varphi  \right\rangle _B},{\left| {{\varphi ^ \bot }} \right\rangle _B}),\\
\end{split}
\end{equation}
where the minimum in the first term is taken over all decompositions $\left\{
{{p_i},\;{{\left| {{\psi _i}} \right\rangle }_{AB}}} \right\}$ and
all basis $\left\{ {{{\left| \psi  \right\rangle }_A},\;{{\left|
{{\psi ^ \bot }} \right\rangle }_A}} \right\}$, and the minimum in the second term is taken over all decompositions $\left\{
{{p_i^{'}},\;{{\left| {{\psi _i^{'}}} \right\rangle }_{AB}}} \right\}$ and
all basis $\left\{ {{{\left|
\varphi  \right\rangle }_B},{{\left| {{\varphi ^ \bot }}
\right\rangle }_B}} \right\}$.
The last two terms are the definition for $LS_A$ and $LS_B$. Thus, $LS
\ge {{(LS_A + LS_B)} \mathord{\left/
 {\vphantom {{(LS_A + LS_B)} 2}} \right.
 \kern-\nulldelimiterspace} 2}$. Q.E.D.

4. Proof for nonzero local superposition in state
\begin{equation}
\label{eq4-5-1}
{\rho _{AB}} = \frac{1}{2}{\left| 1 \right\rangle _A}\left\langle 1 \right| \otimes {\left|  +  \right\rangle _B}\left\langle  +  \right| + \frac{1}{2}{\left|  +  \right\rangle _A}\left\langle  +  \right| \otimes {\left| 1 \right\rangle _B}\left\langle 1 \right|,
\end{equation}
where ${\left|  +  \right\rangle _{A,B}} = \frac{1}{{\sqrt 2 }}\left( {{{\left| 0 \right\rangle }_{A,B}} + {{\left| 1 \right\rangle }_{A,B}}} \right)$.

Assume for this state, $L{S_A} = 0$. According to theorem 2,  ${\rho _{AB}}$ can be written as
\begin{equation}
\label{eq4-50}
{\rho _{AB}} = {p_1}{\left| \varphi  \right\rangle _A}\left\langle \varphi  \right| \otimes \rho _B^1 + {p_2}{\left| {{\varphi ^ \bot }} \right\rangle _A}\left\langle {{\varphi ^ \bot }} \right| \otimes \rho _B^2.
\end{equation}
By tracing part B, that is ${\rho _A} = {\rm Tr}_B({\rho _{AB}})$, ${\rho _A}$ reduces to
\begin{equation}
\label{eq4-5-2}
{\rho _A} = {p_1}{\left| \varphi  \right\rangle _A}\left\langle \varphi  \right| + {p_2}{\left| {{\varphi ^ \bot }} \right\rangle _A}\left\langle {{\varphi ^ \bot }} \right|.
\end{equation}
The above state is the diagonal form of ${\rho _A}$. For the given state ${\rho _{AB}}$ in Eq. (\ref{eq4-50}), by tracing part B,
\[{\rho _A} = \frac{1}{2}{\left| 1 \right\rangle _A}\langle 1| + \frac{1}{2}{\left|  +  \right\rangle _A}\langle  + |.\]
Diagonalizing it and comparing it to Eq. (\ref{eq4-5-2}), we have
\[\begin{array}{l}
{p_1} = \frac{{2 + \sqrt 2 }}{4},\;{p_2} = \frac{{2 - \sqrt 2 }}{4}\\
{\left| \varphi  \right\rangle _A} = \frac{1}{{\sqrt {4 + 2\sqrt 2 } }}\left[ {{{\left| 0 \right\rangle }_A} + (1 + \sqrt 2 ){{\left| 1 \right\rangle }_A}} \right]\\
{\left| {{\varphi ^ \bot }} \right\rangle _A} = \frac{1}{{\sqrt {4 + 2\sqrt 2 } }}\left[ {{{\left| 0 \right\rangle }_A} + (1 - \sqrt 2 ){{\left| 1 \right\rangle }_A}} \right].
\end{array}\]
Thus, ${\left| 1 \right\rangle _A}$ and ${\left|  +  \right\rangle _A}$ can be written as
\[\begin{array}{l}
{\left| 1 \right\rangle _A} = \frac{{\sqrt {2 + \sqrt 2 } }}{2}{\left| \varphi  \right\rangle _A} - \frac{{\sqrt {2 - \sqrt 2 } }}{2}{\left| {{\varphi ^ \bot }} \right\rangle _A}\\
{\left|  +  \right\rangle _A} = \frac{{\sqrt {2 + \sqrt 2 } }}{2}{\left| \varphi  \right\rangle _A} + \frac{{\sqrt {2 - \sqrt 2 } }}{2}{\left| {{\varphi ^ \bot }} \right\rangle _A}.
\end{array}\]
Taking them into Eq. (\ref{eq4-5-1}),
\[\begin{array}{l}
{\rho _{AB}} = \frac{{2 + \sqrt 2 }}{4}{\left| \varphi  \right\rangle _A}\left\langle \varphi  \right| \otimes \frac{1}{2}({\left|  +  \right\rangle _B}\langle  + | + {\left| 1 \right\rangle _B}\langle 1|)\\
\;\;\;\;\;\; + \frac{{2 - \sqrt 2 }}{4}{\left| {{\varphi ^ \bot }} \right\rangle _A}\left\langle {{\varphi ^ \bot }} \right| \otimes \frac{1}{2}({\left|  +  \right\rangle _B}\langle  + | + {\left| 1 \right\rangle _B}\langle 1|)\\
\;\;\;\;\;\; + \frac{{\sqrt 2 }}{4}{\left| \varphi  \right\rangle _A}\left\langle {{\varphi ^ \bot }} \right| \otimes \frac{1}{2}({\left| 1 \right\rangle _B}\langle 1| - {\left|  +  \right\rangle _B}\langle  + |)\\
\;\;\;\;\;\; + \frac{{\sqrt 2 }}{4}{\left| {{\varphi ^ \bot }} \right\rangle _A}\left\langle \varphi  \right| \otimes \frac{1}{2}({\left| 1 \right\rangle _B}\langle 1| - {\left|  +  \right\rangle _B}\langle  + |).
\end{array}\]
Comparing the above equation to the right side of Eq. (\ref{eq4-50}), and noticing that ${\left| \varphi  \right\rangle _A}\left\langle \varphi  \right|, {\left| {{\varphi ^ \bot }} \right\rangle _A}\left\langle {{\varphi ^ \bot }} \right|, {\left| \varphi  \right\rangle _A}\left\langle {{\varphi ^ \bot }} \right|, {\left| {{\varphi ^ \bot }} \right\rangle _A}\left\langle \varphi  \right|$ are linearly independent, to make the right side of these two equations equal, the following equality must hold,
\[\frac{1}{2}({\left| 1 \right\rangle _B}\langle 1| - {\left|  +  \right\rangle _B}\langle  + |)=0.\]
However, the above equality obviously does not hold. Thus, the assumption is false. So, $L{S_A} > 0$. Since $LS
\ge {{(LS_A + LS_B)} \mathord{\left/
 {\vphantom {{(LS_A + LS_B)} 2}} \right.
 \kern-\nulldelimiterspace} 2}$, thus, $LS > 0$ also holds. Q.E.D.
\end{appendix}

\end{document}